\documentclass[12pt]{article}
\usepackage{amssymb}
\usepackage{epsfig}

\newcommand{\beqn}{\begin{eqnarray}}
\newcommand{\eeqn}{\end{eqnarray}}
\newcommand{\beq}{\begin{equation}}
\newcommand{\eeq}{\end{equation}}
\newcommand{\eq}[1]{(\ref{#1})}
\newcommand{\tr}{\mbox{Tr}}

\def\beq{\begin{equation}}
\def\eeq{\end{equation}}
\def\bea{\begin{eqnarray}}
\def\eea{\end{eqnarray}}
\def\bq{\begin{quote}}
\def\eq{\end{quote}}

\def\ba{\begin{array}}
\def\ea{\end{array}}

\begin{document}
\sloppy
 ~ \vspace{-1cm}
\begin{flushright}
{ ITEP-LAT/2004-5}

\vspace{0.2cm}

{ MPI-2004-95}

\end{flushright}

\begin{center}
{\baselineskip=24pt {\Large \bf Three dimensional vacuum  domains in four
dimensional $SU(2)$ gluodynamics}\\

\vspace{1cm}

{\large
    A.\,V.\,Kovalenko$^{\dag}$,
    M.\,I.\,Polikarpov$^{\dag}$,
    S.\,N.\,Syritsyn$^{\dag}$ and
    V.\,I.\,Zakharov$^{*}$} } \vspace{.5cm} {\baselineskip=16pt { \it

$^{\dag}$ Institute of Theoretical and  Experimental Physics,
B.~Cheremushkinskaya~25, Moscow, 117259, Russia\\
$^{*}$ Max-Planck Institut f\"ur Physik, F\"ohringer Ring 6, 80805,
M\"unchen, Germany} }
\end{center}

\vspace{5mm}

\date{}

\abstract{Performing lattice simulations of the four dimensional $SU(2)$
gluodynamics we find evidence for existence of three-dimensional domains whose
total volume scales in physical units. Technically, the domains are defined in
terms of the minimal density of negative links in $Z(2)$ projection of gauge
fields. The volume can be viewed also as the minimal volume bound by the center
vortices. We argue that the three-dimensional domains are closely related to
confinement.}

\newpage

{\bf 1.} It is a general trend in modern theoretical physics to consider
extended objects, like strings and membranes. Usually, one applies these ideas
to hypothetical, high-dimensional completions of the four-dimensional world.
However, lower-dimensional structures might also exist in four dimensions. At
present time there is no well developed theory which would predict such
structures. However, there is accumulating evidence obtained within the lattice
QCD that there are lower dimensions objects percolating through the vacuum of
four dimensional Yang-Mills theories. We have in mind in particular monopoles
and P-vortices, for review see, e.g., \cite{polikarpov,greensite}. The length
of the percolating monopole cluster $L_{perc}$ scales in physical
units~\cite{bornyakov}:
\beq
\label{length} L_{perc}~\approx~(30.8)~(fm)^{-3}\cdot V_4\, ;
\eeq
where $V_4$ is the volume of the lattice.
Similarly, the area of the
P-vortices scales in the physical units~\cite{gubarev}:
\beq\label{area} A_{vort}~\approx~24(fm)^{-2}\cdot V_4~.
\eeq

Scaling laws (\ref{length}), (\ref{area}) have been known since some time
\cite{polikarpov,greensite} but were not originally interpreted
 as an evidence for existence of structures of lower dimensions. The reason is
that mostly monopoles and vortices were thought of  as being `bulky' field
configurations with typical sizes of order $\Lambda_{QCD}^{-1}$. It is only
rather recently that it was recognized that at least at presently available
lattices they do look actually as infinitely thin trajectories and surfaces,
i.e. represent physical structures of lower dimensions. The basic observation
which leads to this conclusion is that the monopoles and vortices are
distinguished by ultraviolet divergent action, see in particular
\cite{gubarev,anatomy}. In view of this there arises a highly non-trivial
question on the consistency of the observations with the asymptotic freedom at
short distances. The data appear to be consistent with the asymptotic freedom
\cite{vz}.

There is no regular way to search for lower-dimensional defects in the vacuum
state of a lattice gluodynamics. Historically, the monopoles and P-vortices are
singled out since they emerged as candidates for confining field
configurations. Both monopoles and P-vortices are defined in terms of projected
fields. In case of the monopoles one uses Maximal Abelian projection while in
case of P-vortices one projects the original $SU(2)$ fields onto the closest
$Z(2)$ gauge field configuration. The use of a projection makes theoretical
analysis on the fundamental level difficult.  Results (\ref{length}),
(\ref{area}) are empirical observations.

In this paper we are looking for possible defects using negative links as the
scanning means. Namely, we use first a $Z(2)$ projection and then minimize the
number of the negative links by the residual $Z(2)$ gauge transformations. The
motivation for such a procedure as well as first encouraging results can be
found in \cite{spz}. In brief, negative links correspond to large potentials,
$A\sim 1/a$ in the continuum limit. Minimizing potentials, on the other hand,
might result in gauge invariant quantities, for a related discussion
see \cite{stodolsky}.

We find, indeed, that the negative links, after minimization,
occupy a part of the lattice which scales as a 3d volume in physical units:
\beq\label{volume}
V_3~\approx~ 2(fm)^{-1}\cdot V_4~.
\eeq

To check the projection (in)dependence we perform measurements both in the
Direct- and Indirect-Maximal Center Projections (DMCP and IMCP). The details of
calculations are given in the Appendix. As a result of $SU(2) \to Z(2)$
projection, the original $SU(2)$ field configurations get projected into the
closest configuration of $Z(2)$ gauge fields. The remaining $Z(2)$ gauge
freedom is then fixed by maximizing the functional
\beq
F(Z) = \sum_{x,\mu} Z_{x,\mu} \label{FZ2}
\eeq
with respect to $Z(2)$ gauge transformations ($Z_{x,\mu} \to z_x\, Z_{x,\mu} \,
z_{x+\hat{\mu}}$, $z_x=\pm 1$). In analogy with the lattice $U(1)$ Landau gauge
fixing the maximization of the functional (\ref{FZ2}) is called $Z(2)$ Landau
gauge fixing.

\begin{figure}[h]
\includegraphics[scale=0.5, angle=270]{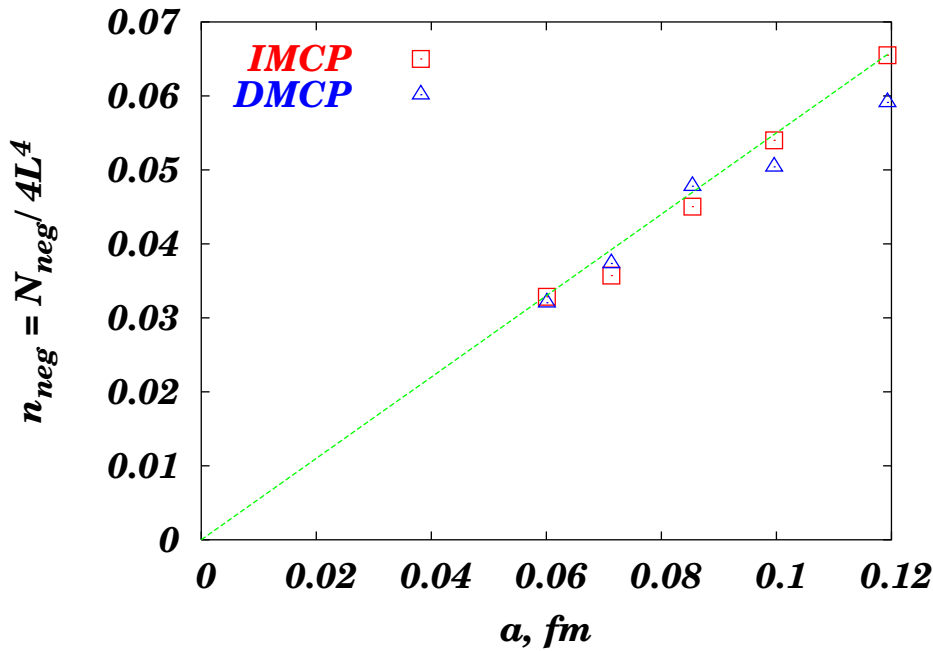}
\caption{Probability for a link to be negative in $Z(2)$ gauge field
configuration after $Z(2)$ Landau gauge fixing.
    \label{fig:neg_links_vs_a}}
\end{figure}

{\bf 2.} Our results for the minimized density of the negative links are shown
in  Fig.~\ref{fig:neg_links_vs_a}. As is seen from the data, both in case of
IMCP and in case of DMCP the density of negative links is proportional to the
lattice spacing.

Let us emphasize that this does imply that there is a strong correlation
between the negative links and they are accumulating within three-dimensional
volumes. Indeed, plaquettes dual to P-vortices do have negative links by
definition. Thus, if we had isolated negative links, then their number would be
less than the number of negative plaquettes (six negative plaquettes correspond
to each isolated link). The data indicate, on the other hand, that the number
of negative links is larger than the number of the negative plaquettes. This
could be interpreted only as indication that there are many plaquettes with two
or four negative links so that they do not contribute to the density of the
P-vortices but do contribute to density of the negative links. The negative
link is dual to a 3d elementary cube, the collection of these cubes form a 3d
volumes which is bound by P-vortices (which also belong to the dual lattice).
The $Z(2)$ Landau gauge corresponds to minimization of the volumes spanned on
$Z(2)$ gauge invariant P-vortices. Thus, it appears that there exists a three
dimensional volume on the dual lattice whose volume scales in the physical
units, the fit of the data of Fig.~\ref{fig:neg_links_vs_a} gives
(\ref{volume}).

{\bf 3.} Let us mention again that the action associated with the vortices and
monopoles is ultraviolet divergent \cite{gubarev, anatomy}:
\beq\label{uv}
S_{vort}~\approx~0.53{A_{vort}\over a^2}~~,~S_{mon}~\approx~\ln 7{L_{perc}\over
a}~,
\eeq
where $S_{vort} = \beta (1 - \frac 12 <\tr U_P^{vort}>)$ is measured on the
plaquettes $U_P^{vort}$ dual to P-vortices. The action $S_{mon} = \beta (1 -
\frac 12 <\tr U_P^{mon}>)$ is measured on the plaquettes nearest to monopole
currents. From the theoretical point of view it is crucial that (\ref{uv}) is
consistent with the asymptotic freedom \cite{vz}.

However the ultraviolet divergence of the action on the plaquettes belonging to
the $3d$ volume would contradict QCD
(for a related discussion see \cite{vz}).
We have measured this action, $S_{3d}$,
and found that, indeed,
\beq\label{zero}
<S_{3d}> ~-~<S>_{lattice}^{without PV}~\approx~0~~,
\eeq
in agreement with theoretical expectations.

{\bf 4.} Now, we will argue that the 3d volumes discussed above might be
relevant to the mechanism of the confinement. Indeed, let us remind the reader
procedure introduced in Ref. \cite{deforcrand}. One determines $Z_{x,\mu}$
variables (see, e.g., the Appendix below) and then replaces the original link
variables $U_{x,\mu}$ by new matrices $\tilde{U}_{x,\mu}$ defined as
\beq\label{modification}
\tilde{U}_{x,\mu}~\equiv~U_{x\mu}\cdot Z_{x,\mu}~~,
\eeq
where the $Z$ factors are $\pm 1$. A remarkable observation of Ref.
\cite{deforcrand}  is that the string tension evaluated by using
$\tilde{U}_{\mu}$ matrices vanishes and there is no confinement for the
modified vacuum.

Usually this procedure is dubbed as `removal' of the P-vortices.
Note, however, that the total area of vortices scales
in physical units and in this sense they represent d=2 defects. Moreover,
the Wilson line is obviously a d=1 subspace. Thus, in d=4
the subspaces d=1 and d=2 do not intersect at all, generally speaking.
For a local change (like (\ref{modification})) to affect the Wilson
line one should change fields  at least on a d=3 subspace.
Thus, no local change of the plaquettes belonging to the vortices can
eliminate confinement. The resolution of the paradox is that the
field modification (\ref{modification}) affects originally links, not
plaquettes directly.

Thus, one can ask what is the minimal number of links which are to be changed
to eliminate confinement through the procedure (\ref{modification}). In this
way we come back to the 3d volume considered in the bulk of the paper.
Moreover, on average the number of intersections of a Wilson line with the d=3
volume is given by (see Fig.~\ref{fig:neg_links_vs_a} and eq.~(\ref{volume})):
\begin{equation}\label{intersection}
\langle N_{intersection}\rangle~\approx~0.5 {P_W\over fm}~~,
\end{equation}
where $P_W$ is the perimeter of the Wilson line. Note that the number of the
intersections (\ref{intersection}) does not depend on the lattice spacing $a$.

It is worth emphasizing that if one chooses randomly links where the sign of
the matrix $U_{x,\mu}$ is changed then the area law cannot be affected and only
extra dependence of the Wilson loop on the perimeter could be generated. Thus,
elimination of the confinement through multiplication of the Wilson loop by
$(-1)^{N_{intersection}}$ is possible only in case of a coherent effect. In
this context, one can claim these 3d defects to be responsible for the
confinement.

Of course the above discussion does not rule interpretation a la Bohm-Aharonov
effect (which is related to the linking number of the surface with magnetic
flux and the Wilson line in 4d space~\cite{ABfield}). Indeed, it is not
possible to eliminate 3d volumes without eliminating their boundary, the
P-vortices. Moreover the modified Wilson loop $W'=\tr\prod_{l\in C}\{U_l'\}$
can be expressed through the initial Wilson loop, W, as:
\beq
W'~=~\exp\{i \pi{\cal L(P,C)}\}~W\, ,
\eeq
here $\cal L$ is the 4d linking number of the surface of the P-vortex, $\cal
P$, and the Wilson contour $\cal C$. $\cal L$ can be expressed as the number of
intersections of $\cal C$ with the 3d volumes, or as the integral over $\cal P$
and $\cal C$.

Finally let us note that recently 3d manifolds corresponding to the topological
charge density of the definite sign were found~\cite{horvath} in 4d lattice
$SU(2)$ gauge theory.

{\bf 5.} The authors are grateful to J. Greensite and F.V.~Gubarev for
illuminating discussions. A.V.K., M.I.P. and S.N.S. are partially supported by
grants RFBR 02-02-17308, RFBR 01-02-17456, DFG-RFBR 436 RUS 113/739/0, RFBR-DFG
03-02-04016, INTAS-00-00111, and CRDF award RPI-2364-MO-02. V.I.Z. is partially
supported by grant INTAS-00-00111  and by DFG grant ``From lattice to hadron
phenomenology''.

\section*{Appendix}

We perform our calculations both in the Direct~\cite{greens2} -- and the
Indirect~\cite{debbio} -- Maximal Center Projections (DMCP and IMCP). The DMCP
in SU(2) lattice gauge theory is defined by maximization of the functional
\begin{equation}
F_1(U) = \sum_{x,\mu} \left( \tr U_{x,\mu}\right)^2 \, , \label{maxfunc}
\end{equation}
with respect to gauge transformations, $U_{x,\mu}$ is the lattice gauge field.
Maximization of (\ref{maxfunc}) fixes the gauge up to Z(2) gauge
transformations and the corresponding Z(2) gauge field is defined as:
$Z_{x,\mu} = \mbox{sign} \tr U_{x,\mu}$. To get IMCP we first fix the maximally
Abelian gauge maximizing the functional
\begin{equation}
F_2(U) = \sum_{x,\mu} \tr \left( U_{x,\mu}\sigma_3 U_{x,\mu}^+ \sigma_3\right)
\, , \label{maxfuncmaa}
\end{equation}
with respect to gauge transformations. We project gauge degrees of freedom
$U(1)\to Z(2)$ by the procedure completely analogous to the DMCP case, that is
we maximize $F_1(U)$ (\ref{maxfunc}) with respect to $U(1)$ gauge
transformations.

To fix the maximally Abelian and direct maximal center gauge we create 20
randomly gauge transformed copies of the gauge field configuration and apply
the Simulated Annealing algorithm to fix gauges. We use in calculations that
copy which correspond to the maximal value of the gauge fixing functional. To
fix the indirect maximal center gauge from configuration fixed to maximally
Abelian gauge and to fix the Z(2) degrees of freedom one gauge copy is enough
to work with our accuracy. We work at various lattice spacings to check the
existence of the continuum limit. The parameters of our gauge field
configurations are listed in Table~\ref{conf_table}. To fix the physical scale
we use the string tension in lattice units~\cite{Fingberg:1992ju}, $\sqrt\sigma
= 440\, MeV$.

\begin{table}[t]\label{conf_table}
\caption{Parameters of configurations.}
\begin{center}
\begin{tabular}{|c|c|c|c|}
\hline
$\beta$ & Size &    $N_{IMCP}$ &    $N_{DMCP}$ \\
\hline
$2.35$ &    $16^4$ &    $20$ &  $20$ \\
$2.40$ &    $24^4$ &    $50$ &  $20$ \\
$2.45$ &    $24^4$ &    $20$ &  $20$ \\
$2.50$ &    $24^4$ &    $50$ &  $20$ \\
$2.55$ &    $28^4$ &    $37$ &  $17$ \\
$2.60$ &    $28^4$ &    $50$ &  $20$ \\
\hline
\end{tabular}
\end{center}
\end{table}

\end{document}